# Evaluating the retrieval effectiveness of Web search engines using a representative query sample

Dirk Lewandowski
Hamburg University of Applied Sciences, Department of Information, Finkenau 35, D—22081 Hamburg, Germany
dirk.lewandowski@haw-hamburg.de

## ABSTRACT

Search engine retrieval effectiveness studies are usually small-scale, using only limited query samples. Furthermore, queries are selected by the researchers. We address these issues by taking a random representative sample of 1,000 informational and 1,000 navigational queries from a major German search engine and comparing Google's and Bing's results based on this sample. Jurors were found through crowdsourcing, data was collected using specialised software, the Relevance Assessment Tool (RAT). We found that while Google outperforms Bing in both query types, the difference in the performance for informational queries was rather low. However, for navigational queries, Google found the correct answer in 95.3% of cases whereas Bing only found the correct answer 76.6% of the time. We conclude that search engine performance on navigational queries is of great importance, as users in this case can clearly identify queries that have returned correct results. So, performance on this query type may contribute to explaining user satisfaction with search engines.

## INTRODUCTION

Given the importance of search engines in everyday use (Purcell, Brenner, & Raine, 2012; van Eimeren & Frees, 2012) which is also expressed by the many millions of queries entered into general-purpose Web search engines every day ("comScore Reports Global Search Market Growth of 46 Percent in 2009," 2010), it is surprising to find how low the number of research articles on search engine quality still is. One might assume that large-scale initiatives to monitor the quality of results from general-purpose search engines such as Google and Bing already exist. Users could refer to such studies to decide which search engine is most capable of retrieving the information they are looking for. Yet as we demonstrate below, existing search engine retrieval effectiveness studies have covered a much smaller sample so far. This small size has cast a degree of doubt on their conclusions.

One could argue that it is the search engine vendors' sole responsibility to conduct tests on the retrieval effectiveness of their engines in relation to their competitors (and they surely conduct such studies). The goal of studies conducted by the vendors and academic studies respectively, however, are quite different. While the goal of studies conducted by search engine vendors is to improve a certain search engine (and test additional capabilities such as dealing with search engine spam), the purpose of academic studies is to advance retrieval effectiveness testing and provide a valid picture of the quality of different search engines. Decision-making, revolving for instance around funding alternative search engines, is facilitated with this information.

Although search engines are using click-through data to a large degree to improve their services and existing research focuses on this area (Chuklin & Serdyukov, 2013; Joachims et al., 2007; Joachims, Granka, Pan, Hembrooke, & Gay, 2005), search engine providers still conduct classic retrieval effectiveness tests using human jurors recruited for that task (*Google Search Quality Rating Guidelines*, 2012).

Retrieval effectiveness tests are just one part of the overall evaluation of search engine quality (Lewandowski & Höchstötter, 2008). Measuring the effectiveness of search engine results is an integral part of such frameworks and is usually based on a TREC-style setting. However, these methods have been criticised as being inadequate for evaluating Web search engines, where results presentation differs from other information retrieval systems (Lewandowski, 2013a). Furthermore, retrieval effectiveness tests have been criticized for not taking into account actual user behaviour, i.e., they are based on the query-response paradigm and do not consider interaction in the searching process (Ingwersen & Järvelin, 2005). While these criticisms should not be ignored, continued enhancement of query-response-based methods is still a valuable pursuit.

Many search engine retrieval effectiveness tests have been conducted in the past. One problem they all share is that they only use a relatively low number of queries. The selection of queries is also not representative of the queries entered into actual search engines. In this paper, we present a study using 1,000 queries representatively selected from the transaction logs of a major German search portal. To our knowledge, this is the first attempt to use a representative query sample for a search engine retrieval effectiveness study. The general goal of this paper is to show how large-scale search engine retrieval effectiveness studies can be conducted, and to present valid results based on a representative query sample.
The rest of this article is structured as follows: First, we will review related work on search engine retrieval effectiveness. Here we will focus on testing for different query types, the general design of such studies, and the selection process for queries in such tests. Next we will present our objectives and research questions, followed by our methods and findings. We then discuss our findings and summarize them in the conclusions section, explaining limitations and providing some suggestions for further research.

## LITERATURE REVIEW

A lot of work has already been done on measuring search engine retrieval effectiveness. Table 1 provides an overview of the major search engine retrieval effectiveness studies published over the last two decades. In this literature review, we focus on different query types used in search engine retrieval effectiveness studies, on the general design of such tests, and on the selection of queries to be used.

### Tests on different query types

It has been found that search engines are used for a very wide range of purposes. A good distinction between query intents (a.k.a., query types) was presented by Broder (2002). This author differentiates between

- Informational queries, where the user aims at finding some documents on a topic he/she is interested in
- Navigational queries, where the user aims at navigating to a Web page already known or where the user at least assumes that a specific Web page exists. A different expression for such queries is “homepage finding task” (as used in TREC). As Hawking and Craswell (2005, p. 219) put it, “The judging criterion is, is this the page I wanted? - that is, the home page of the entity I was thinking of or the page I named?”
- Transactional queries, where the user wants to find a Web page where a further transaction (e.g., downloading software, playing a game) can take place.

While there have been attempts to improve this classification (mainly through adding sub-categories), for most purposes, the basic version still allows for a sufficiently differentiated view (for a further discussion on query intents and their usage in search engines, see Lewandowski, Drechsler, & Mach (2012).
The vast majority of search engine retrieval effectiveness studies focus on informational queries (see table 1, and next section). Studies on navigational queries are rather rare, and it should be stressed that as the information need is satisfied with just one result, different measures should be used when evaluating search engines on this query type — see Hawking & Craswell (2005) and Lewandowski (2011a). Measures used in such studies include the number of queries answered/unanswered, success N, and the mean reciprocal rank.
There are also some “mixed studies” such as Griesbaum (2004) that use query sets consisting of informational as well as navigational queries. These studies are flawed because, in contrast to informational queries, the expected result of a navigational query is just one result. When more results are considered, even a search engine that found the desired page and placed it on the first position would receive lower precision scores. Therefore, informational and navigational queries should be separated.
Regarding transactional queries, we found no empirical studies. This may have to do with the fact that transactional queries are difficult to identify. Often, it is not easy to distinguish between navigational and transactional intent. Furthermore, in some cases, there is only one relevant result for such a query (e.g., a Website identified through the query, where the user wants to play a specified game), and in other cases, there are multiple relevant results (e.g., a user wants to play a specified game that is available on multiple Websites). We still lack a model for evaluating search engine performance for transactional queries.

### Retrieval effectiveness study design

In general, retrieval effectiveness tests rely on methods derived from the “classic” information retrieval test designs (Tague-Sutcliffe, 1992) and on advice that focuses more on Web search engines (Gordon & Pathak, 1999; Hawking, Craswell, Bailey, & Griffiths, 2001). A sample of queries is taken and then sent to the information retrieval systems under investigation. The results are collected and mixed randomly to avoid learning and branding effects. Jurors then judge on the relevance of the results. Then, results are once again allocated to the search engines, and results are analysed. Search engine retrieval effectiveness studies have adopted this basic design but adjusted it to the specifics of search engines.

Most investigations only consider the first 10 or 20 results. This has to do with the general behaviour of search engine users. These users only rarely view more than the first results page (Pan et al., 2007; Spink & Jansen, 2004). Whereas in the past, these results pages typically consisted of a list of 10 results (the infamous “ten blue links”), search engines now tend to add additional results (e.g., from news or video databases), so that the total amount of results especially on the first results page is often considerably higher (Höchstötter & Lewandowski, 2009). Furthermore, researchers have found that general Web search engine users heavily focus on the first few results (Cutrell & Guan, 2007; Granka, Hembrooke, & Gay, 2006; Hotchkiss, 2007; Joachims et al., 2005; Pan et al., 2007) which are shown “above the fold” (Höchstötter & Lewandowski, 2009) — i.e., the part of the first search engine results page (SERP) that is visible without scrolling down. Keeping these findings in mind, a cut-off value of the first 10 results in search engine retrieval effectiveness tests seems reasonable.

An important question is how the results should be judged. Most studies use relevance scales with three to six points or binary relevance decisions. Using binary relevance judgments is straightforward, as many different types of calculations (arithmetic mean, etc.) can be used on this type of data. However, they only have a limited expressiveness (Lewandowski, 2013b), and it remains unclear where the threshold between relevant and irrelevant exactly lies (Greisdorf & Spink, 2001). Other studies (Ding & Marchionini, 1996; Eastman & Jansen, 2003; Véronis, 2006) have used graded relevance judgements, but there has been no agreement on what scale to use. Griesbaum (Griesbaum, Rittberger, & Bekavac, 2002; Griesbaum, 2004) uses binary relevance judgments with one exception: results can also be judged as “pointing to a relevant document” (i.e., the page itself is not relevant but has a hyperlink to a relevant page). This is done to take into account the special nature of the Web, where it is easy for a user to follow a link from an irrelevant result pointing to a relevant document. However, it seems problematic to judge these pages as (somehow) relevant, as pages can have many links, and a user then (in the worst case scenario) has to follow a large number of links to finally access the relevant document.

An important point in the evaluation of search engine results is that the source of results is made anonymous, i.e., the jurors do not know which search engine delivered a certain result). Otherwise, they would prefer results from their favourite search engine. Studies found that users mostly prefer Google’s results, even when they are displayed in the page layout of another search engine, and vice versa (Jansen, Zhang, & Schultz, 2009).

It is also important that the result lists are randomized, due to possible learning effects. While most of the newer studies do make the source of the results anonymous (as to which search engine produced the results), we found only five studies that randomize the results lists (Gordon & Pathak, 1999; Hawking, Craswell, Thistlewaite, & Harman, 1999; Jansen & Molina, 2006; Lewandowski, 2008; Véronis, 2006).

Most studies use students as jurors. This comes as no surprise, as most researchers teach courses where they have access to students who can serve as jurors. In some cases (again, mainly older studies), the researchers themselves judge the documents (Chu & Rosenthal, 1996; Ding & Marchionini, 1996; Dresel et al., 2001; Griesbaum et al., 2002; Griesbaum, 2004; Leighton & Srivastava, 1999). To our knowledge, there is no search engine retrieval effectiveness study that uses more than one juror per search result. This could lead to flawed results because different jurors may not agree on the relevance of individual results, and they may also assign different information needs to the same task or query (see Huffman & Hochster, 2007). However, using more than one juror per result is often not feasible in practice.

A trend we have observed is that methods for retrieval effectiveness studies are being applied in other contexts, e.g., for testing search engines which are designed to be used by children (Bilal, 2012). Another trend involves adjusting retrieval effectiveness studies to the results presentation in search engines, which is not simply list-based anymore but instead combines different elements like organic results, contextual advertising (which should be seen as a result type), and results from vertical search indexes injected into the main results lists as part of what is known as “universal search”, often involving highlighting (Lewandowski, 2013a).

There are also studies which either compare search engines to other information systems (e.g., question answering systems (Vakkari, 2011) or use a combination of methods for measuring retrieval effectiveness and measurements of correlations between relevance judgments and other criteria, e.g., correlating relevance judgments for Wikipedia articles found in search engines with an assessment using lexical criteria (Lewandowski & Spree, 2011), or correlating relevance judgments with commercial intent of the documents found (Lewandowski, 2011b).
There is also a trend towards carrying out more international and country-specific studies — see Lazarinis, Vilares, Tait, & Efthimiadis (2009) for an overview.

### Query selection in search engine retrieval effectiveness studies

In an ideal setting, a search engine retrieval effectiveness study would use a representative sample of queries. Obtaining data from commercial search engine transaction logs and conducting a large-scale study are however difficult due to the amount of data to be collected and the number of relevance judgments to be made by jurors. Consequently, the body of research on the topic consists only of relatively small-scale studies, and the number of queries used in the studies varies greatly. Particularly the oldest studies (Chu & Rosenthal, 1996; Ding & Marchionini, 1996; Leighton & Srivastava, 1999) use only a few queries (between 5 and 15) and are, therefore, of limited use. Newer studies use at least 25 queries, and some use 50 or more. For a discussion of the minimum number of queries that should be used in such tests, see Buckley & Voorhees (2000). We can see that this number is far from being representative, and it often remains unclear which criteria are used to select queries.
In older studies, queries are usually taken from reference questions or commercial online systems, while newer studies focus more on the general users' interest or mix both types of questions (Can, Nuray, & Sevdik, 2004; Lewandowski, 2008; Tawileh, Griesbaum, & Mandl, 2010). There are studies that deal with a special set of query topics such as business (Gordon & Pathak, 1999), e-commerce (Jansen & Molina, 2006), or library and information science (Deka & Lahkar, 2010). While this approach would improve the external validity of the studies (if queries are at least representative for the field studied), due to the low number of queries used and the selection process applied, studies using general-purpose queries are not representative. Using lists with popular queries provided by major search engines is not a good idea either, as these lists are heavily filtered (Sullivan, 2006), e.g., to remove adult content queries.
As we can see from table 1, only a relatively low number of queries was selected for each study. Queries are manually selected, and therefore the external validity of the studies is questionable. However, these studies are still useful, as they often do not seek to answer the ultimate question, "Which search engine is the best?", but rather to experiment with improving the methods for retrieval effectiveness studies. For example, Griesbaum (2004) and Lewandowski (2008) added results descriptions ("snippets") to the evaluation framework, and Lewandowski (2011a) proposed methods for evaluating search engines on their performance on navigational queries (see below). However, to our knowledge there is no study that collected data for both at the same point in time.

## OBJECTIVES AND RESEARCH QUESTIONS

The objective of this research was to conduct a large-scale search engine retrieval effectiveness study using a representative query sample, distributing relevance judgments across a large number of jurors. The goal was to improve methods for search engine retrieval effectiveness studies.
The research questions are:

RQ1: Which search engine performs best when considering the first 10 results?

RQ2: Is there a difference between the search engines when considering informational queries and navigational queries, respectively?

RQ3: Is there a need to use graded judgments in search engine retrieval effectiveness studies?

## METHODS

Our study will also focus on informational and navigational queries. Transactional queries have been omitted because they present a different kind of evaluation problem (see above), since it is not possible to judge on the relevance of a certain amount of documents found for the query, but instead, one would need to evaluate the searching process from query to transaction.

## Study on navigational queries

As a user expects a certain result when entering a navigational query, it is of foremost importance for a search engine to show that exact result at the top position. While search engines may improve their performance on that type of query when further results are considered (Lewandowski, 2011a), users will most often perceive that the search engine has failed if the desired result is not presented on the first position. Therefore, for evaluation purposes, we collected relevance judgements only for the first result presented by each engine, even though in some cases, users surely would profit from a choice of more results. As queries are cannot definitely be classified into one of Broder's categories, but one could rather measure the probability of a certain query belonging to one of the categories (Lewandowski, Drechsler & von Mach, 2012), search engines present more than one result even for queries that are navigational with high confidence. Even for such queries, there may be more than one relevant result. However, for the purpose of this study, we defined navigational queries as ones where a clear distinction between one single correct result and all other irrelevant results can be made. Therefore, there is also no need to use differentiated scales.

We collected the first result for a sample of 1,000 queries (see below) from Google and Bing. A research assistant classified the results that were found either as correct or incorrect, not knowing which search engine provided which results.

## Study on informational queries

For the informational queries, we designed a conventional retrieval effectiveness test, similar to the ones described in the literature review. We used a sample of queries (see below), sent these to the German-language interfaces of the search engines Google and Bing, and collected the top 10 URLs from the results lists. We crawled all the URLs and recorded them in our database. We then used jurors, who were given all the results for one query in random order and were asked to judge the relevance of the results using a binary distinction (yes/no) and a 5-point Likert scale. Jurors were not able to see which search engine produced a given result, and duplicates (i.e., results presented by both search engines) were presented for judgement only once.

## Building the sample

Our aim was to build a sample representative of the queries as entered by the German search engine users. As it was not feasible to get data from all relevant search engine providers, we used data from just one search portal, T-Online, the largest German Web portal. It offers a search engine that uses results from Google and has a market share of 2.9% (measured in number of queries) in Germany (Webhits, 2013). While the market share seems rather low, T-Online is number four in the German search engine market, and as the market is quite large, the total number of queries entered into this search engine is many millions per month. Therefore, while searcher behaviour may surely differ from this one search engine to the others, we think that sampling from T-Online allows for sufficient validity. Our basis was all queries entered within a multi-week period in 2011.[1] The distribution of queries is quite uneven, which means that there is a small ratio of queries that are entered very often, while a large amount of queries are entered only rarely (long tail searches). These characteristics are consistent with previous data as reported by Jansen & Spink (2006). With such a distribution, it is not sufficient to randomly select the sample, as the sample will be flawed due to criteria with many instances not being selected. In our case, for instance, the random selection may only return queries with low or average frequencies, while "important" queries (i.e. higher-volume queries) may be neglected. This problem occurs with all "informetric" distributions (Bookstein, 1990; Stock, 2006), and was for example also discussed by Lawrence & Giles (1999) when considering which hosts to sample for estimating the size of the World Wide Web.

To build our sample, we sorted the queries according to their popularity, and then divided the query list into ten segments of equal size including duplicates. Therefore, the first segment containing the very popular queries consists of only 16 distinct queries (which were entered 3,049,764 times in total). The last segment (also containing 3,049,764 query instances) consists of 643,393 distinct queries, because in the long tail, each query was only entered a few or even a single time in the time-frame considered (see table 2 for details).

---

[1] For competitive reasons, we are not allowed to give the exact time-span here, as it would be possible to calculate the exact number of queries entered into the T-Online search engine within this time period.

From each segment, we selected 360 queries, and classified them as informational, navigational, or other. Only informational and navigational queries were considered for further analysis. Our goal was to have 100 queries of each type in each segment. However, this was not achievable for the first segments, as these did not contain enough distinct queries, due to their high popularity. If the 360 queries classified resulted in more than 100 queries classified as informational and navigational, respectively, we took out queries randomly until that the sample for that segment consisted of 100 queries.
This process lead to a sample that is not just random in that every unique query has the chance to be in the sample, but also considers the distribution of queries, and therefore does not over- or under-represent popular or unpopular queries, respectively.

Table 2: Dividing the query database into segments; sample sizes

| Segment | Number of query instances (cumulative) | Number of distinct queries |
|---|---|---|
| 1 | 3,049,764 | 16 |
| 2 | 6,099,528 | 134 |
| 3 | 9,149,293 | 687 |
| 4 | 12,199,057 | 3,028 |
| 5 | 15,248,821 | 10,989 |
| 6 | 18,298,585 | 33,197 |
| 7 | 21,348,349 | 85,311 |
| 8 | 24,398,114 | 189,544 |
| 9 | 27,447,878 | 365,608 |
| 10 | 30,497,642 | 643,393 |
| | | |

### Collecting search results

We used the *Relevance Assessment Tool* (hereafter referred to as RAT), a self-developed Web-based software (Lewandowski & Sünkler, 2013) to collect the search engine results. The software has a screen-scraping module, which allows for a list of queries to be uploaded. These queries are then automatically sent to the search engines, and the results' URLs are collected. Each URL gets crawled, and a copy of the resulting document gets collected into a database for further analysis (see below). The software is an effective means of data collection, allowing large numbers of results to be collected with ease. We used two query lists with 1,000 navigational and informational queries each.
For the *navigational queries*, we collected only the first result from each search engine, while for *informational queries*, we collected the first ten results. The cut-off value for the latter seems more than reasonable because search engine users in most cases only consider the first results page (usually displaying 10 organic results). The first few results displayed in particular receive a great deal of attention (see above). As there is only one relevant result for a navigational query, we assume that a search engine is only successful when it displays this desired result in the first position. Lewandowski (2011a) found that, while there are gains when more than the first result is considered, in most instances search engines are quite capable of displaying the desired result in the first position, if they manage to find it in the first place.

### Collecting relevance judgments

For collecting the relevance judgments, we also used the RAT (Lewandowski & Sünkler, 2013). Another module of this software allows researchers to distribute tasks to a large number of jurors. In this study, we used a single access code, which can easily be distributed through e-mail or on social networks. Jurors can log in into the tool and then work on a task. The drawback of this approach is that researchers have no control over who the jurors logging in to the system are.
We paid jurors for their participation. For each task completed, a juror was sent an Amazon voucher worth 5€. The RAT generated the e-mails sending the vouchers automatically. We adjusted the tool to generate a voucher when over 90% of the results for a task were graded, keeping in mind that the tool allows jurors to skip some results if they either are not displayed (due to technical difficulties in crawling) or if a juror does not feel competent to judge an individual result.
While we did not track who our jurors were, we distributed the access code mainly through a mailing list reaching library and information science students at our university, and also distributed the code on the department's Facebook page. Jurors were permitted to complete multiple tasks. From the logs where jurors entered their e-mail addresses to get the voucher, we can see that we had an uneven distribution of tasks per juror, where some jurors completed many tasks (i.e., around 15), while many users just completed one or two tasks. Completing a task took about 20 minutes, and through the distributed approach, we were able to collect the relevance judgments for all 1,000 tasks within seven hours.
For the *navigational* queries, a research assistant judged each result as either being the desired Website or not. As this kind of judgment is simple and unambiguous, it is reasonable to have a single person judging all results. For the *informational* queries, we asked the jurors to judge each result twice: Once in a binary way (whether the result was relevant or not), and once on a five-point scale ranging from 0 to 4, where 0 represents completely irrelevant, and 4 means completely relevant. Jurors were allowed to skip results when they were not displayed or in cases in which they felt unable to provide a judgement. Together with results not correctly crawled, this yielded 9,265 and 9,790 graded results for Google and Bing, respectively. Since jurors were allowed not only to skip all possible judgments for a result but also to skip just one particular judgement (e.g., giving a binary relevance judgment, but not a graded one), the number of judgments per result in the different analyses below may differ.

## RESULTS

In this section, we first present the results for the navigational queries and then the results for the informational queries. The latter section comprises the analysis of the binary and graded relevance judgments.

### Navigational queries

As illustrated in figure 1, Google outperformed Bing on the navigational queries. While Google found the correct page for 95.27% of all queries, Bing only did so 76.57% of the time.

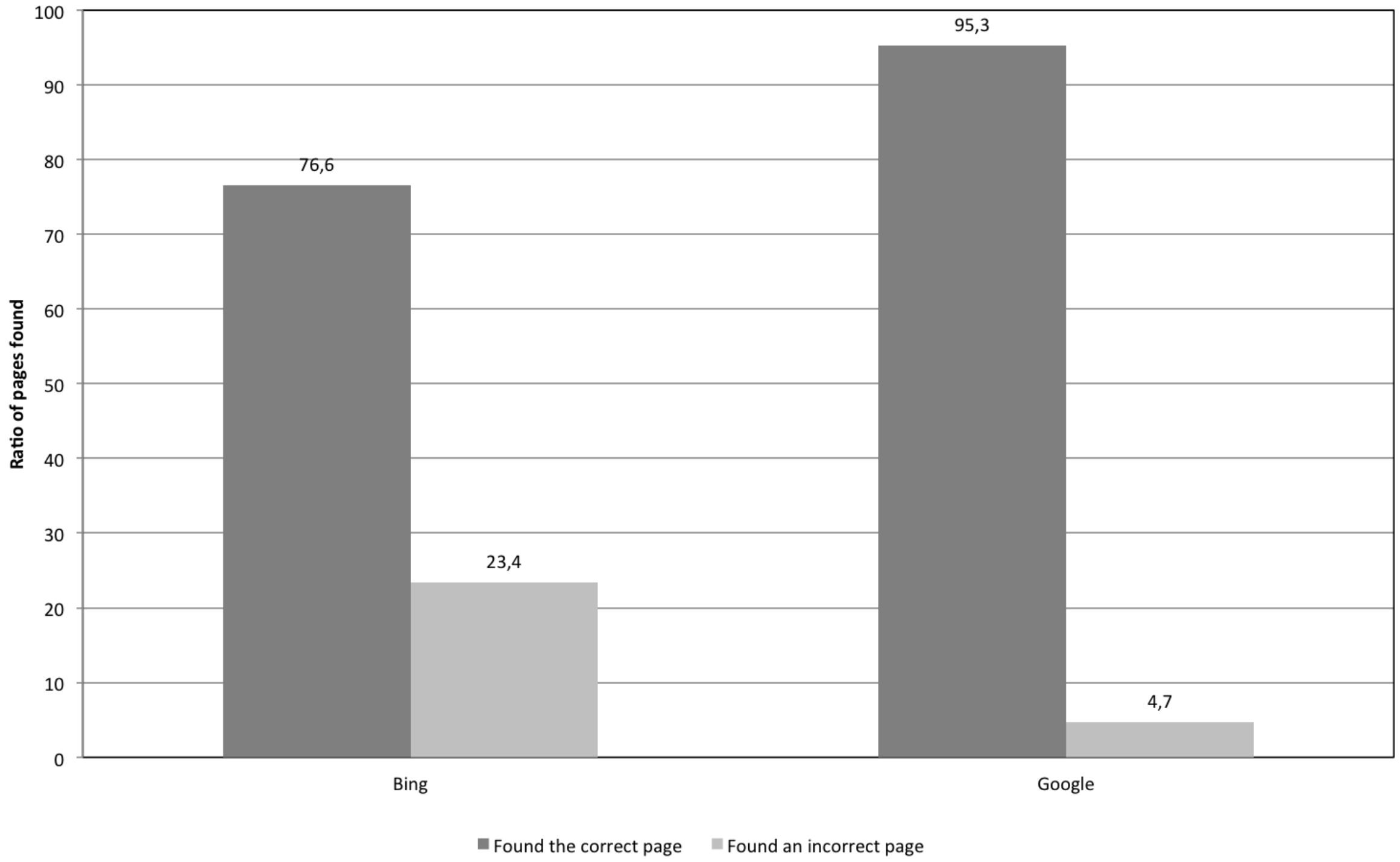

Figure 1. Ratio of relevant vs. irrelevant pages found at the top position (navigational queries)

**Informational queries**

The total ratio of relevant results for the informational queries is 79.3% for Bing, and 82.0% for Google, respectively (figure 2). We can see that Google produces more relevant results than its competitor.

When considering the results ranking (figure 2), we can see that the precision on the first two results positions is comparable between Google and Bing, but when considering more positions, the differences between the search engines increase (figure 3). In the figure, the ratio of relevant results is given for each results position. E.g., when considering position 3, the precision score for all results presented on the first three positions is calculated.

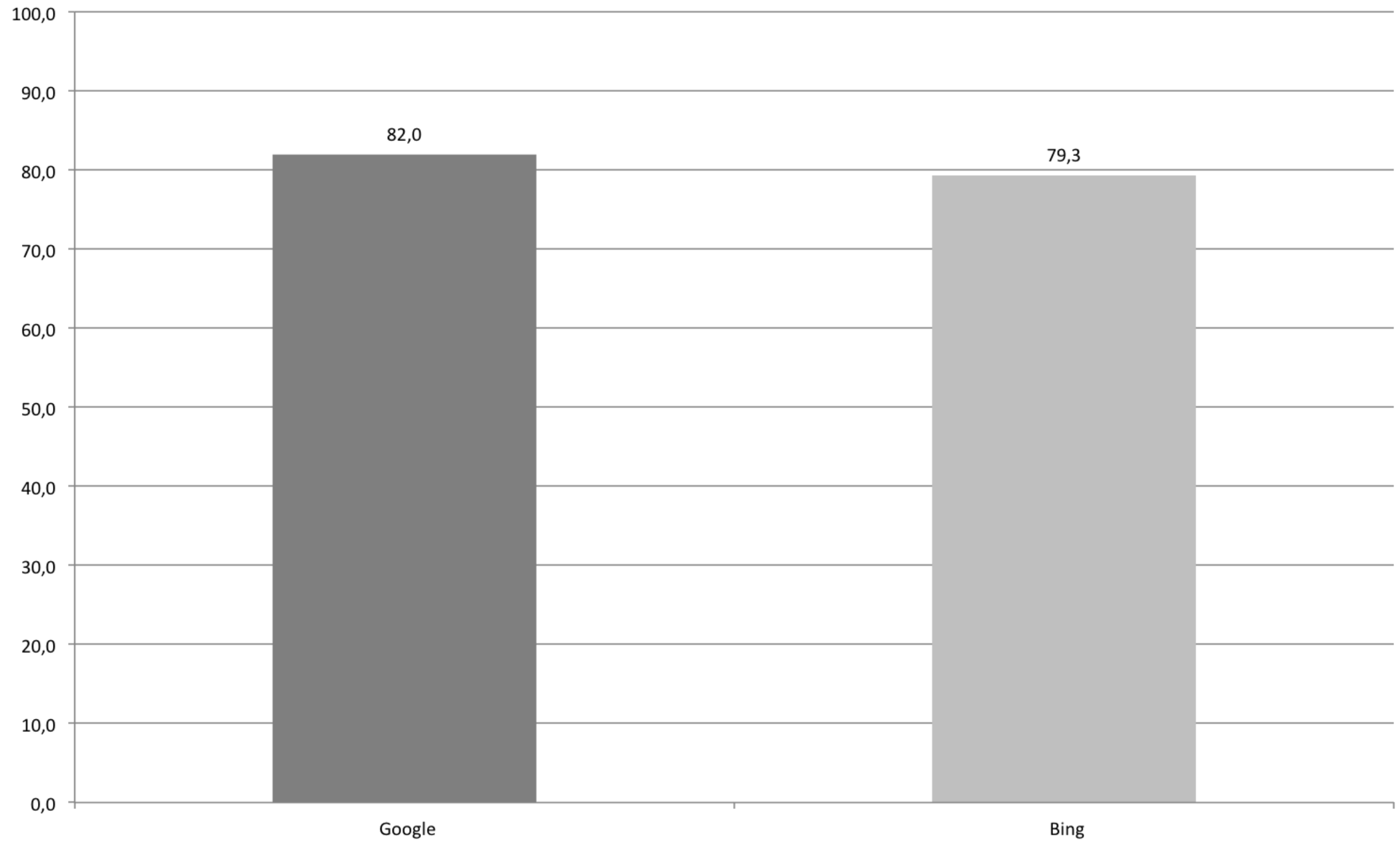

Figure 2. Ratio of relevant vs. irrelevant documents found in the top 10 results (informational queries)

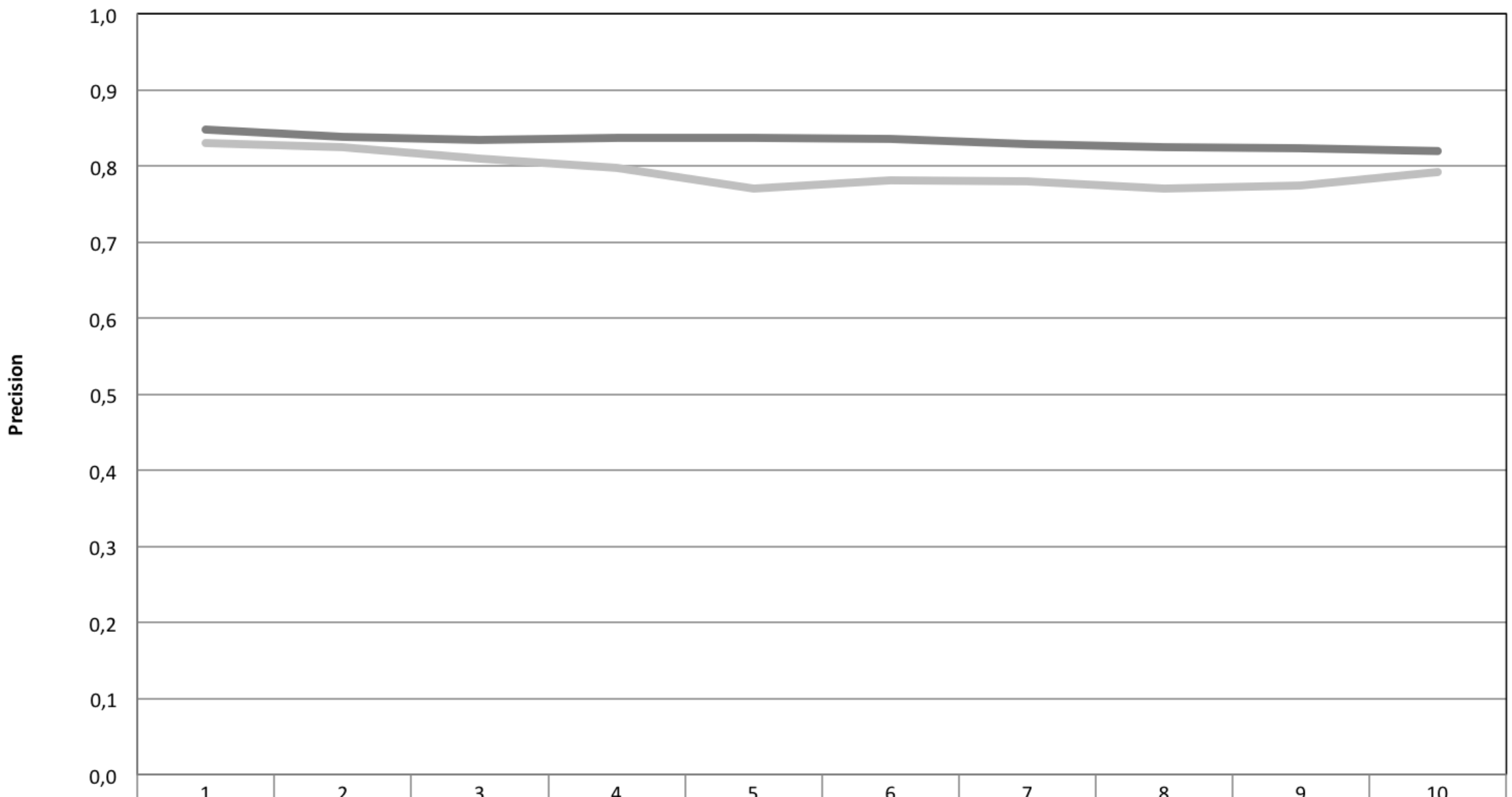

| | 1 | 2 | 3 | 4 | 5 | 6 | 7 | 8 | 9 | 10 |
|---|---|---|---|---|---|---|---|---|---|---|
| Google | 0,847660501 | 0,838274933 | 0,834545455 | 0,836950624 | 0,837527115 | 0,835532261 | 0,829204088 | 0,824770146 | 0,822718004 | 0,819535888 |
| Bing | 0,82970297 | 0,824649299 | 0,809183673 | 0,797775531 | 0,770020534 | 0,781059063 | 0,779608651 | 0,770103093 | 0,77445932 | 0,791534392 |

Position (cumulative)

Figure 3. Recall-precision graph (top 10 results)

Figure 4 shows averages across all judgments according to the results position. Again, the differences are not pronounced. Only when considering the first five results positions does Google’s clear advantage over Bing become apparent. However, when considering all 10 results positions, Bing outperforms Google slightly.

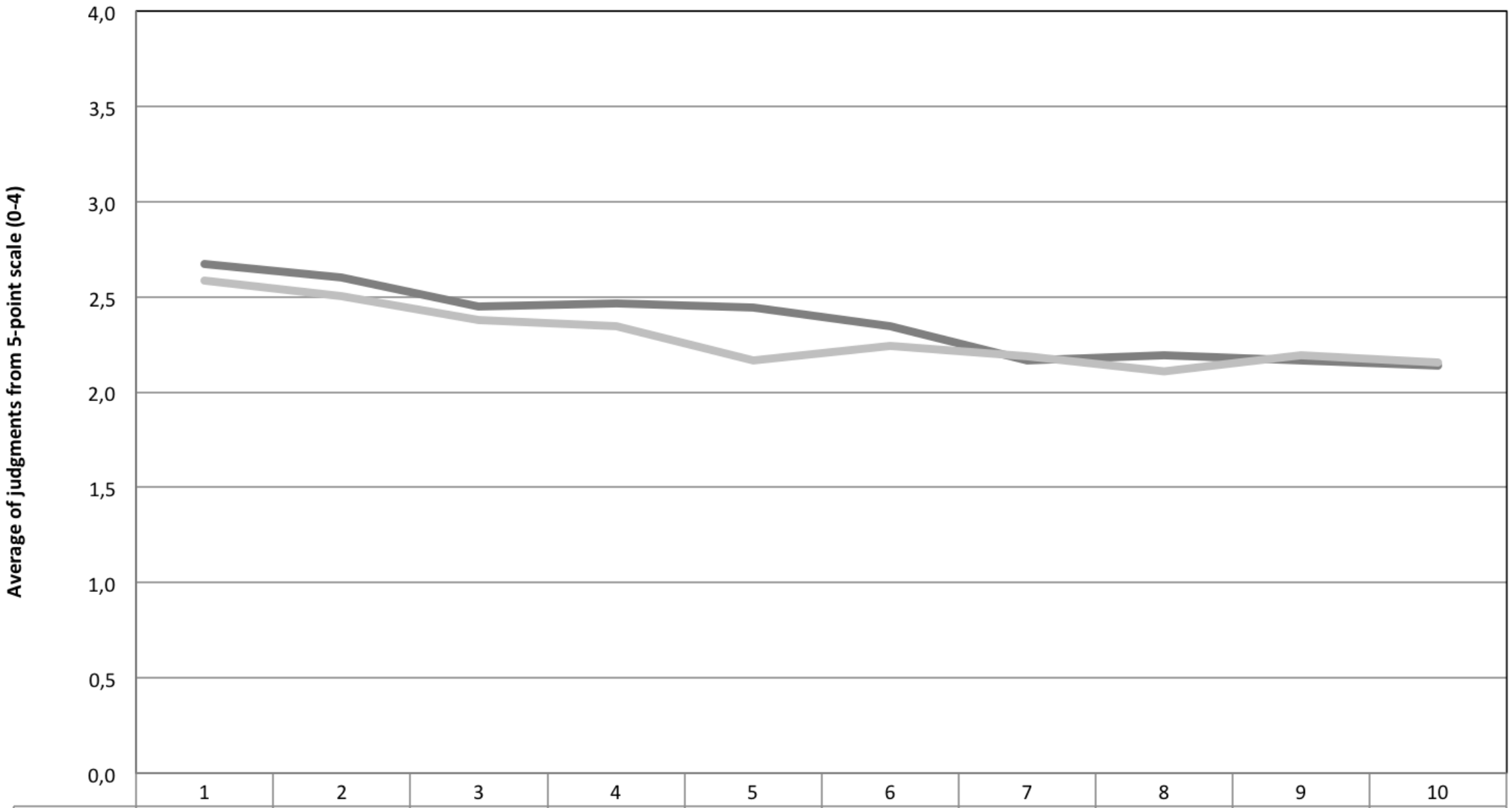

| | 1 | 2 | 3 | 4 | 5 | 6 | 7 | 8 | 9 | 10 |
|---|---|---|---|---|---|---|---|---|---|---|
| Google | 2,672703752 | 2,600512164 | 2,449197861 | 2,465982028 | 2,442748092 | 2,346886912 | 2,166666667 | 2,192405063 | 2,164348925 | 2,137179487 |
| Bing | 2,584705882 | 2,501779359 | 2,379975875 | 2,348139256 | 2,164835165 | 2,242718447 | 2,189189189 | 2,109202454 | 2,194581281 | 2,158291457 |

Figure 4. Average relevance judgments on the Likert scale

Figure 5 shows the ratio of results given different grades of relevance for the two search engines. It can be seen that Google outperforms Bing, but the differences are not excessive. A remarkable difference, however, is that Bing does not produce more of the less relevant results (graded 1 or 2), but rather the difference between Bing and Google is mainly a result of Bing producing a larger number of results judged as completely irrelevant.

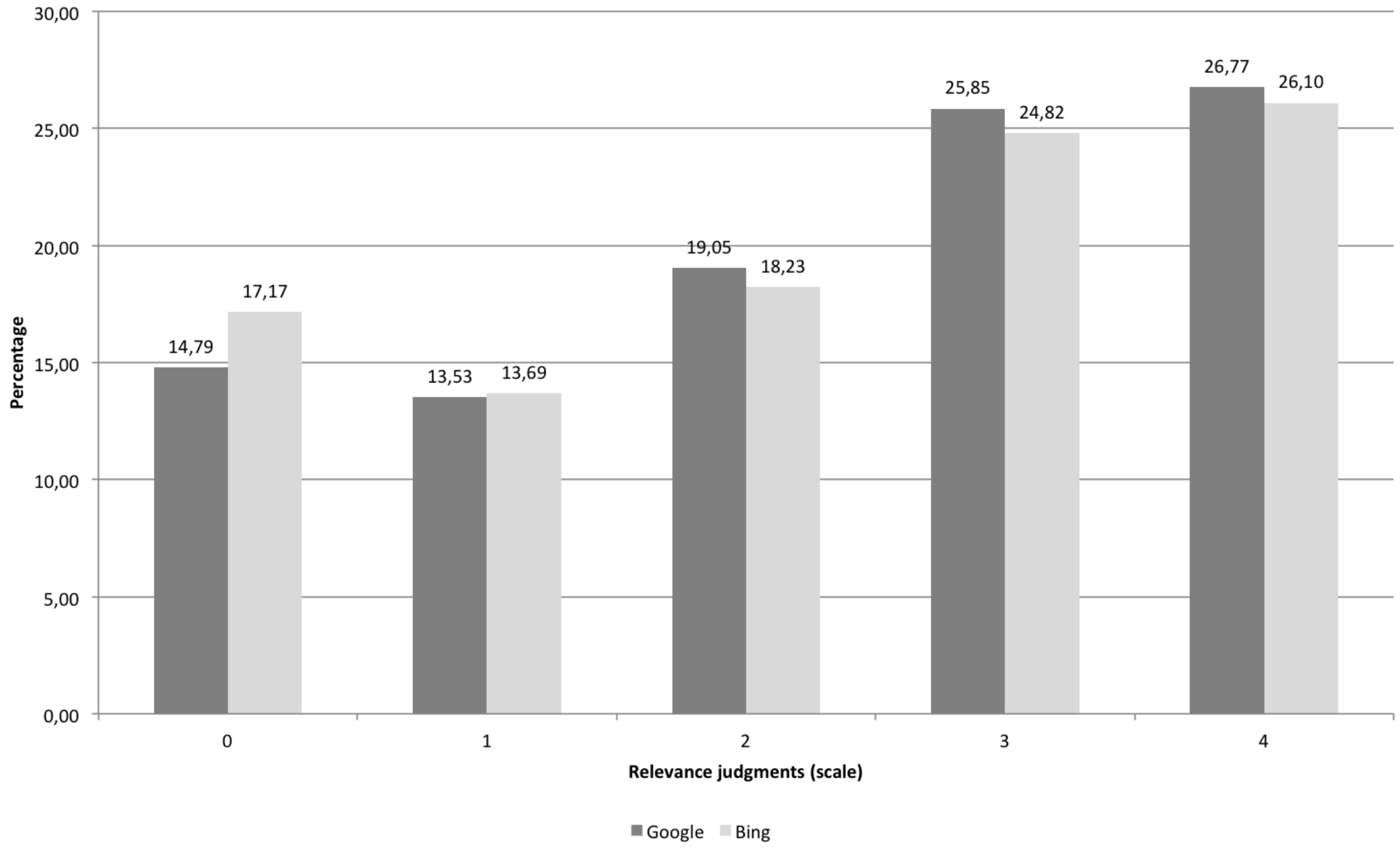

Figure 5. Distribution of the Likert-scale relevance judgments for both search engines

## DISCUSSION

The main difference between the relevance of the results produced by the two search engines in this study is that Google outperforms Bing on the informational queries, and even more so on the navigational queries. As navigational queries account for a large ratio of the total queries – between 15% and 40%, dependent on the source (Lewandowski et al., 2012) – we can reason that the user satisfaction with Google can in part be explained by Google’s good performance on such queries. Users are definitely able to judge the relevance of results for navigational queries, as there is only one correct result for such queries. For informational queries, not only is there more than one relevant result in most cases, it is also hard for users to determine whether or not the results set presented by the search engine is indeed the best possible results. It may well be that (1) one result or even numerous relevant results are missing in the results set of the search engine in use, or (2) there may be equally relevant results presented by another search engine, that contribute aspects not considered in the results presented by the search engine in use. The latter case is especially important since two different search engines could produce results that are of equal relevance, but completely different from one another. Studies examining the overlap of results sets of the major search engines (Spink, Jansen, Blakely, & Koshman, 2006) found that the overlap in the top 10 results was quite low. However, this does not necessarily mean that one search engine produces more relevant results than the others. It could also mean that there are enough relevant results for most queries, so that two completely distinct results sets may be of equal relevance. While this is surely a field for further research, it leads us to the conclusion that further aspects of the results should be considered to determine which search engine produces the best results and to advanced search engine comparisons methodologically. Promising aspects include the credibility of the results (Kammerer & Gerjets, 2012; Lewandowski, 2012), the diversity of the results (Denecke, 2012) and the commercial influence in the results lists (Lewandowski, 2011b).

When looking at the distribution of relevance judgments on the Likert scale, the most remarkable finding from the study is that Google outperforms Bing mainly because it produces significantly fewer results graded as completely irrelevant. We can therefore assume that Bing’s problem does not lie in the fact that it produces fewer relevant results than Google, but rather in identifying whether results are at least somehow relevant to a user. This can be seen more as a spam detection problem than as a ranking problem.

## CONCLUSION

To summarize our findings, we can conclude that Google outperforms Bing overall (RQ1).
With respect to RQ2, there are clear differences regarding the navigational queries, where Google outperforms Bing significantly. While there are also differences regarding the informational queries, these are not too pronounced, and we argue that the differences will not be noticeable to the general user, but only when scientifically comparing search engines.
Regarding the use of graded relevance judgments (RQ3), we argue that while the overall outcome when comparing the two search engines is not different, graded relevance judgments allow one to detect additional effects (i.e., Bing producing a considerably larger amount of completely irrelevant results). We therefore recommend using graded instead of binary relevance judgments.
The results from our study are comparable to previous studies. However, due to the improved sample size, the validity of our study is much greater. The software tool and the crowdsourcing approach make much larger studies possible than before. We encourage researchers to improve the validity of their work through larger studies and better query sampling.
When considering that there is only one correct result for navigational queries, whereas for informational queries, there may be multiple results of equal relevance (and different users may find different results relevant), we argue that Google's superiority as perceived by the search engine users can (at least in part) be explained by its performance on navigational queries. This explanation adds to the literature, where aspects beyond results quality, such as search engine brand (Jansen et al., 2009), are considered when discussing user satisfaction with the results of their favourite search engine.

### Limitations

This study is not without limitations. Firstly, jurors were only given the queries derived from the query logs, but no description of an underlying information need. They therefore had to interpret the query themselves. While this may have led to some misinterpretations, it would not have distorted the results, as a juror always judged the relevance of results from both search engines. We are aware that it is rather unusual only to give the jurors the "naked" queries, but we think it is a better — and less biased way — than for the researchers to create artificial intents from the queries from the log files. We think that constructing information needs (or information need descriptions, respectively) always leads to a bias, while letting the juror interpret the queries at least distributes the bias. A solution for that problem would be to have multiple persons creating information need descriptions from the queries, as done by Huffman & Hochster (2007). However, such an approach does not appear practical when dealing with a large number of queries.
Not knowing the query intent could also have resulted in jurors being too generous with their judgments. However, this bias would also have been seen in both search engines, and thus would not have caused the overall results of the comparison between the engines to be biased. It should be noted, though, that the good results for both search engines (precision scores of around 0.8) may be in part explained by this fact.
Another limitation is that we only used one juror per query. While it is desirable to have as many jurors per query as possible, this is not practical in most cases. We found no major search engine retrieval effectiveness study using more than one juror for the same document.
While we used a representative query sample, this sample can only be representative for the search engine we used. It is not clear to what degree querying behaviour differs between users of different search engines. To our knowledge, there is only one study in which data from different search engines was collected over the same period of time (Höchstötter & Koch, 2009). However, differences in the distribution of queries were not analysed in detail.
A limitation that could have resulted in bias in our study is that errors did occur during data collection. Some results on the Google SERPs did not directly refer to the result's URL, but instead to a Google URL for tracking purposes. The RAT was sometimes not capable of extracting the target URLs from the longer Google URLs, and some of these URLs were therefore not considered valid. Consequently, these results were omitted, and the total number of Google results in the study is considerably lower than the number of Bing results.
Furthermore, we did not collect Universal Search results, i.e., results from vertical search injected into the general results lists. Such results can lead to higher results precision, and the results in our study could be biased due to the fact that presentation of vertical results could lead to the search engines not repeating the same result in the regular results list. If certain highly relevant results were presented in the Universal Search results, they would have been omitted from our study. It is unclear to what extent this would lead to a bias in the results (Lewandowski, 2013a), as both search engines under investigation present Universal Search results, although presumably to different extents.
Both the latter points can be overcome with further development of the software we used for data collection, and we have already taken the first steps in accomplishing this.

### Further research

While our study dealt with some of the methodological problems with retrieval effectiveness testing, we still lack studies addressing *all* the major problems with such tests. In further research, we hope to integrate realistic search tasks, use queries of all three types, consider results descriptions and results as well (Lewandowski, 2008), and examine all results presented on the search engine results pages (Lewandowski, 2013a) instead of only organic results.
We also aim to study the overlap between the relevant results of both search engines beyond simple URL overlap counts (as used, among others, in Spink et al. (2006)). Are the results presented by different search engines basically the same, or is it valuable for a user to use another search engine to get a “second opinion”? However, it should be noted that the need for more results from a different search engine only occurs for a minority of queries. There is no need for more results on already satisfied navigational queries or even informational queries when a user just wants some basic information (e.g., the information need is satisfied with reading a Wikipedia entry). However, we still lack a full understanding of information needs and user satisfaction when it comes to Web searching. Further research should seek to clarify this issue.
At the end of this article, we want to stress why search engine retrieval effectiveness studies remain important. There is still no agreed-upon methodology for such tests. While in other contexts, test collections are used, the problem with evaluating Web search engines in that way is that (1) results from a test collection cannot necessarily be extrapolated to the whole Web as a database (and no test collection even approaches the size of commercial search engines’ index sizes), and (2) the main search engine vendors (i.e., Google and Bing) do not take part in such evaluations.
Secondly, there is a societal interest in comparing the quality of search engines. Search engines have become a major means of knowledge acquisition. It follows that society has an interest in knowing their strengths and weaknesses.
Thirdly, regarding the market share of specific Web search engines (with Google’s market share at over 90% in most European countries, see Lunapark (2011)), it is of interest whether the dominance of just one player is a result of its superior quality, or whether other factors need to be considered to explain the situation.

Table 1: Search engine retrieval effectiveness studies (expanded from (Lewandowski, 2008).

| Authors | Year | Query language | Number of queries | Number of results | Number of engines tested | Engines tested | Query topics | Jurors | Information need stated? | Relevance judged by actual user? | Relevance scale | Source of results made anonymous? | Results lists randomised? |
|---|---|---|---|---|---|---|---|---|---|---|---|---|---|
| Chu & Rosenthal (Chu & Rosenthal, 1996) | 1996 | English | 10 | 10 | 3 | AltaVista, Excite, Lycos | Queries drawn from real reference questions | Persons conducting the test | no | no | scale (0; 0.5; 1) | no | no |
| Ding & Marchionini (Ding & Marchionini, 1996) | 1996 | English | 5 | 20 | 3 | Infoseek, Lycos, OpenText | 3 randomly selected from a question set for Dialog online searching exercises in an information science class, 2 personal interest | Person conducting the test | no | no | scale (6 point) | no | no |
| Leighton & Srivastava (Leighton & Srivastava, 1999) | 1999 | English | 15 | 20 | 5 | AltaVista, Excite, HotBot, Infoseek, Lycos | Reference queries at a University library | Person conducting the test | yes | no | categories (4) + duplicate link + inactive link | yes | |
| Gordon & Pathak (Gordon & Pathak, 1999) | 1999 | English | 33 | 20 | 8 | AltaVista, Excite, Infoseek, OpenText, HotBot, Lycos, Magellan, Yahoo Directory | Business-related | Faculty from a business school | yes | yes | scale (4 point) | no | yes |
| Hawking & Craswell (Hawking et al., 1999) | 1999 | English | 54 | 20 | 7 | Northern Light, Snap, AltaVista, HotBot, Microsoft, Infoseek, Google, Yahoo, Excite, Lycos, Euroseek, MetaCrawler, Inquirus, FAST, EuroFerret, DirectHit, ANZwers, ExciteAus, Web Wombat, LookSmart Australia | general interest | Australian with university degree | no | no | binary | yes | yes |
| Wolff (Wolff, 2000) | 2000 | German | 41 | 30 | 4 | AltaVista, C4, MetaCrawler, NorthernLight | 20 general interest, 21 scientific | | no | yes | scale (3: rel, non rel, maybe rel) | yes | ? |
| Dresel et al. (Dresel et al., 2001) | 2001 | German | 25 | 25 | 6 | Abacho, Acoon, Fireball, Lycos, Web.de directory, Yahoo directory | five groups: products, guides, scientific, news, multimedia | Person conducting the test | no | no | binary | no | no |
| Griesbaum et al. (Griesbaum et al., 2002) | 2002 | German | 56 | 20 | 4 | Google, Lycos, AltaVista, Fireball | scientific; society | students and faculty | no | no | binary/scale (3 point) | yes | no |
| Eastman & Jansen (Eastman & Jansen, 2003) | 2003 | English | 100 | 10[2] | 3 | AOL, Google, MSN | Queries with operators (AND, OR, MUST APPEAR, and PHRASE) | | no | no | scale (4 point) | no | no |
| Can, Nuray & Sevdik (Can et al., 2004) | 2003 | English | 25 | 20 | 8 | AlltheWeb, AltaVista, HotBot, InfoSeek, Lycos, MSN, Netscape, Yahoo | general interest | Students and professors | yes | yes | binary | yes | ? |
| Wu & Li (Wu & Li, 2004) | 2004 | English | 24 | 20 | 8 | Google, AllTheWeb, HotBot, AltaVista, MetaCrawler, ProFusion, MetaFind, MetaEUREKA | From TREC 2002 Web topic destillation task | | no | no | Binary | ? | no |
| Griesbaum (Griesbaum, 2004) | 2004 | German | 50 | 20 | 3 | Google, Lycos, AltaVista | general interest | students | no | no | binary/scale (3 point) | yes | no |
| Jansen & Molina (Jansen & Molina, 2006) | 2005 | English | 100 | 10[3] | 5 | Excite, Froogle, Google, Overture, Yahoo! Directory | ecommerce queries | College students | no | no | scale (3 point) | yes | yes |
| Véronis (Véronis, 2006) | 2006 | French | 70 | 10 | 6 | Google, Yahoo, MSN, Exalead, Dir, Voila | 14 topic areas; general interest | students | no | yes | scale (0-5) | yes | yes |
| MacFarlane (MacFarlane, 2007) | 2007[4] | English | 50 | 10 | 4 | Google, AltaVista, Lycos, Yahoo | general interest | n/a | n/a | no | Binary | n/a | n/a |
| Deka & Lahkar | 2007 | English | 50 | 10 | 5 | Google, Yahoo!, Live, Ask, AOL | library and information science | n/a | no | no | | no | no |
| Lewandowski (Lewandowski, 2008) | 2008 | German | 40 | 20 | 5 | Google, Yahoo, MSN, Ask, Seekport | general interest | students | yes | yes | binary, scale (1-5), scale (0-100) | yes | yes |
| Tawileh, Mandl & Griesbaum (Tawileh et al., 2010) | 2009 | Arabic | 50 | 10 | 5 | Araby, Ayna, Google, MSN, Yahoo! | Random set of top ten queries of Arabic search engine | Independent jurors (students and engineers) | yes | no | Binary | yes | no |

[2] Minus duplicates within first ten search results

[3] Plus sponsored links

[4] Data collected in 2001